\documentstyle[12pt,aps,pre,psfig]{revtex}

\newcommand {\be}{\begin{equation}}
\newcommand {\ee}{\end{equation}}
\newcommand {\bea}{\begin{eqnarray}}
\newcommand {\eea}{\end{eqnarray}}

\begin{document}

\title{Estimation of Lyapunov spectra from space-time data}
\author{Martin J.\ B\"unner \footnote{published in Phys. Lett. A
{\bf 258} (1999) 25.}, R. Hegger $^\dagger$}
\address{Istituto Nazionale di Ottica\\ Largo E.\ Fermi,
I-50125 Firenze, Italy \\ $^\dagger$ MPIPKS, N\"othnitzer Str. 38, D-01087 Dresden, Germany}

\date{\today} \maketitle

\begin{abstract}
A method to estimate Lyapunov spectra from spatio-temporal
data is presented, which is well-suited to be applied to
experimental situations.  It allows to characterize the
high-dimensional chaotic states, with possibly a large number of
positive Lyapunov exponents, observed in spatio-temporal chaos. The
method is applied to data from a coupled map lattice.
\end{abstract}
\vspace{2cm}

keywords: nonlinear time series analysis, spatio-temporal chaos

\vspace{2cm} Nonlinear spatially-extended systems (SES) attracted a
vast research activity in nonlinear dynamics, partly because of their
ability to exhibit high-dimensional chaotic motion, with possibly a
large number of positive Lyapunov exponents (for an overview on SES
see \cite{Cross93}). Established dynamical models for SES are partial
differential equations (PDE) for systems continuous in space and time,
and coupled map lattices (CML) for systems discrete in space and
time. The spectrum of Lyapunov exponents can be directly estimated
from the models by numerical means.  On the basis of the Lyapunov
spectrum the Kaplan-Yorke dimension and the metric entropy can be
estimated.  Along these lines, an extensive scaling of the Lyapunov
spectrum, as well as the Kaplan-Yorke dimension and the metric
entropy, with respect to the system size has been observed for several
PDE- and CML-models \cite{Ruelle82}.  The term 'spatio-temporal chaos'
has been coined to describe this phenomenology. Additionally, various
well-controlled experiments to investigate the nature of
spatio-temporal chaos were conducted. Unfortunately, it was not
possible to directly extract the Lyapunov spectrum of the experimental
systems from the space-time data so far, since one has to find an
appropriate nonlinear model first. Typically, this is done by using a
time-delayed embedding (of high enough order) to construct a phase
space from a scalar measurement \cite{Kantz97}. Then, the
time-dependent Jacobian can be estimated from the data. Unfortunately,
one faces two problems for relatively high dimensions of phase space
(i. e. $D \geq 6$): (1) An exponentially increasing number of data is
required. (2) The delay embedding induces a folding in phase space,
which makes it increasingly difficult to perceive a deterministic
structure \cite{Olbrich97}. For the above reasons the established
techniques for the estimation of Lyapunov spectra are restricted to
chaotic states with a small number of positive Lyapunov exponents (say
$N^{+}$ smaller or equal to 3). Besides that spatio-temporal
chaos was characterized by the decomposition into spatial modes 
such as the Karhunen-Loeve decomposition \cite{Politi90,Zoldi97}.

In this paper we address the question of how to estimate the spectrum
of Lyapunov exponents of a SES from space-time data, which could
i.e. be the output of an experiment. Typically, the equation of
motion is unknown. Recently, novel methods for modelling
SES on the basis of models with a purely local coupling (PDE, CML)
were established \cite{Voss98,Parlitz98,Baer98}.
We show how the latter local models  can be used to
estimate the Lyapunov spectrum of a SES from space-time data.  Then,
we apply the method to space-time data of a CML and compare the
results to the theoretical prediction. Finally, we show how the method
can be used to investigate boundary effects and the 'universal' nature
of the Lyapunov spectrum of a homogeneous SES.

Let us recall the definition of the spectrum of Lyapunov exponents of
a non-autonomous dynamical system 
\be 
\vec{x}_{n+1} =
\vec{f}(\vec{x}_n,\vec{y}_n),
\label{map}
\ee
where $\vec{x}_n \in {\cal R}^I, \vec{y}_n \in {\cal R}^J$ is 
an external driving force, and $\vec{f}: {\cal R}^{I+J} \to {\cal R}^I$. 
For the ease of presentation we restrict the following discussion on time-discrete systems.
The spectrum of Lyapunov exponents $\lambda_i, i=1,...,I$ 
of the system (\ref{map}) is defined as 
the contraction rates of a volume element averaged over
the attractor \cite{Kantz97,Casdagli92}:
\be 
\lambda_i (\vec{x}_0, \vec{y}_n) = \lim_{T \to \infty} \frac{1}{T} \ln |\Lambda_i^{(T)}|,
\label{lyapdef}
\ee
where $\Lambda_i^{(T)}$ are the eigenvalues of the product of all Jacobians 
${\bf J}_n = \frac{\delta \vec{f}}{\delta \vec{x}}(\vec{x}_n,\vec{y}_n)$:
\be 
\prod_{n=1}^T {\bf J}_n  \vec{u}_j^{(T)} = \Lambda_i^{(T)}  \vec{u}_i^{(T)}.
\label{ew}
\ee In general, the spectrum depends on the initial state $\vec{x}_0$,
and the sequence of inputs $\vec{y}_n$.  In special cases, these
dependences can be removed with the help of multiplicative ergodic
theorems. It has been shown that in the autonomous case
($\vec{y}_n$=0), as well as for random inputs $\vec{y}_n$ the limit
(\ref{lyapdef}) exists and only depends on the invariant measure 
\cite{Kantz97,Casdagli92}.
In basically all relevant cases the Lyapunov spectrum $\lambda_i$
cannot be computed from the equations (\ref{map})- (\ref{ew}) directly
and one has to rely on numerical estimations of the Lyapunov spectrum.
A well-established technique for its estimation from time-evolution
equations (\ref{map}) is described in \cite{Kantz97}: The trajectory
as well as $I$ perturbation vectors are iterated numerically (with the
help of the time-dependent Jacobian ${\bf J}_n$).  A successive
orthonormalization of the perturbation vectors prevents them from
alignment.
From the Lyapunov spectrum important invariant quantities can be estimated: 
1) The Kaplan-Yorke dimension,   
$D_{KY}= k + \frac{\sum_{i=1}^k \lambda_i}{|\lambda_{k+1}|}$, where
$\sum_{i=1}^k \lambda_i \geq 0$ and $\sum_{i=1}^{k+1} \lambda_i < 0$. 
2) An upper bound for the metric entropy via the Pesin formula,  $h=\sum_{i=1}^{N^+}\lambda_i$, where $N^+$ is the number of positive Lyapunov exponents.

At first we concentrate for simplicity on discrete, scalar systems, with a local
 coupling only,   
which are modeled with the help of coupled map lattices (CML).  The
cases of continuous dynamics in space and time, as well as
multi-component systems can be treated analogously and will be
discussed at the end of the paper. Let the experimental data
$\{x_n^i\}$ be a measurement of a homogeneous, scalar
spatially-extended system with a discrete spatial variable $i=1,...,I$
and a discrete time $n=1,..n=1,..,N$. The overall number of data is
$NI$. If the unknown equation of motion of the system under
investigation is a CML with a local coupling of range $m$, the data
can be modeled by 
\be x^i_{n+1} = \hat{h}(\vec{v}_n^i),
\label{CML}
\ee with $\vec{v}_n^i=(x_n^{i-m},\ldots,x_n^{i+m})$ and $\hat{h}:
{\cal R}^{2m+1} \to {\cal R}$, for $i=m+1,..,I-m$.  The system
(\ref{CML}) is non-autonoumous, since it is driven by the variables at
the boundaries $(x_n^1,\ldots, x_n^m,x_n^{I-m},\ldots,x_n^I)$. The
phase space of such a system has $(N-2m)$ dimensions, allowing for a
high-dimensional chaotic motion with possibly a lot of positive
Lyapunov exponents. Nevertheless, the restricted couplings of the
variables in phase space - as it is expressed by the function
$\hat{h}$, which couples only four values of the space-time data -
allows to obtain a nonlinear, deterministic model in a
$(2m+2)$-dimensional space only.
It is of fundamental importance to statistically verify the model class (\ref{CML}) 
on the basis of the data with appropriate tools, 
which also allow to estimate an appropriate coupling range $m$.
Since this has been described in detail elsewhere, we refer to 
the literature \cite{Voss98,Parlitz98,Baer98}.

In this paper we use a local linear model $\hat{h}(\vec{v}_n^i) =
\vec{a}_n^i \vec{v}_n^i + b_n^i$, where the parameters $(\vec{a}_n^i, b_n^i)$ 
can be obtained for each time $n, n=2,...,N$ and each position $i,i=m+1,...,I-m$ by the 
minimisation of the distance 
\be
\sigma_n^i =\sqrt{ \frac{1}{N_{U_n^i}} \sum_{\vec{v}_n^i \in U_n^i} 
\bigg( x^i_{n+1} -  \hat{h}(\vec{v}_n^i) \bigg)^2 },
\label{min}
\ee of the data to the local linear model with the help of a local
least squares fit. $U_n^i$ is a sufficiently small neighbourhood of
$\vec{v}_n^i$ with the cardinal number $N_{U_n^i}$. Since the model
$\hat{h}$ is homogeneous, appropriate neighbors can be chosen with
respect to a varying time $n$ as well as a varying position $i$. This
allows to treat every space-time point $x_n^i$ (except at the
boundaries) on the same basis and to vastly enhance the statistics and
decrease the requirements on the number of data by averaging over
space and time.  The local linear fits have been chosen for practical
purposes; we note that in principle any appropriate method for nonlinear fitting
can be used.

As a consequence of the local coupling, 
the Jacobian ${\bf J}_n(x_n^1,\ldots,x_n^{m},x_n^{I-m},\ldots,x_n^I)$ 
of a SES  has the form of a $(2m+1)$-diagonal  matrix, 
which in the case of a CML-model (\ref{CML}) reads,
\be
 {\bf J}_n^{(ij)} = \left\{
	\begin{array}{lll}
	  0, & & \text{if $|i-j| > m$},\\
	  \frac{\partial \hat{h}}{\partial x_n^j}(\vec{v}_n^i), & & \text{else,}
	\end{array} \right. 
\label{Jacobian}
\ee for $i,j=m+1,..,I-m$. In the case of a local linear fit for
$\hat{h}$, the non-zero entries of the Jacobian (\ref{Jacobian}) are
given by the coefficients $\vec{a}_n^i$.  Let us recall that the
spectrum of Lyapunov exponents (\ref{lyapdef}) depends on the sequence
of inputs, which in the case of SES are the boundaries.  While in
priniciple, this dependence on the sequence of inputs at the
boundaries cannot be removed, we found some evidence that if the
investigated system is part of a larger system, which displays
homogeneous spatio-temporal chaos, the Lyapunov exponents do not
depend on the sequence of inputs at the boundaries.  This will be
discussed in more detail later.
 
We performed several tests of the above presented method for established models of SES. At first, we
present results on the estimation of Lyapunov spectra for a lattice of coupled logistic maps
\be
x^i_{n+1} = (1-2\epsilon) f(x_n^{i}) + \epsilon (f(x_n^{i-1}) + f(x_n^{i+1})),
\label{log}
\ee
with $f(x)=4x(1-x)$,
from space-time data $\{x_n^i\}, n=1,...,10,000, i=1,...,50$ 
with fixed boundary conditions $x_n^1=x_n^I=0.1$ and $\epsilon=0.2$. 
In this case, since the boundary conditions are fixed, the system is
autonomous with a 48-dimensional phase space.  
The first step is to identify the data $\{x_n^i\}$ to be governed
by a local model (\ref{CML}) and to estimate the appropriate value
of the coupling range $m$. Since the estimation and verification 
of local models from space-time data has been already described in detail,
we refer to the literature \cite{Voss98,Parlitz98,Baer98}, and
from now on consider the model (\ref{CML}) to be verified.
The function $\hat{h}$
was modelled with a local linear model under varying $m$, the
neighborhoods $U^i_n$ were chosen such that at least 30 points were
contained. The 
coefficients of the Jacobian ${\bf J}_n$ were extracted according to
(\ref{Jacobian}).  We propagated $(50-2m)$ perturbation vectors in
time according to the fitted Jacobian ${\bf J}_n$ and orthonormalized
the perturbation vectors after each time step.  The spectrum of the
Lyapunov exponents was computed as a mean over the logarithms of the
stretching factors of the propagation vectors (for details see
\cite{Kantz97}).  We compare the results to the estimation of the
Lyapunov spectrum from the equations (\ref{log})
As presented in Fig. 1. we find a good agreement of the spectra for
$m \ge 1$. The estimation of the exponents for $m=0$
does not yield a meaningful result and therefore the $(m=0)$-model has
to be rejected. Since the Lyapunov spectrum under variation of the
coupling range $m$ converges for $m \ge 1$, we propose the convergence
of the LS as another criterion for estimating $m$.
In Table 1 we compare the dimension, the entropy and the maximum Lyapunov
for $m=0,1,2$. 

Our approach of a non-autonomous model (\ref{CML}) allows to estimate
Lyapunov spectra not only in the case in which all variables
$i=1,...,I$ are observed, but also for incomplete observations in
space $i=i_0,...,i_0+l-1$. The so-defined subsystems of length $l$ are
driven by the input sequences at the boundaries
($x_n^{i_0},\ldots,x_n^{i_0+m} ,x_n^{i_0+l-m}, \ldots,x_n^{i_0+l-1}$),
where in general the Lyapunov spectrum also depends on. Therefore, the
dimension of the phase space is $l-2m$ and the dimension density
$\delta$ is calculated as $\delta = \frac{D_{KY}}{l-2m}$, and the
entropy density $\eta$ is calculated as $\eta=
\frac{H}{l-2m}$.

At first, we present results of the estimation of the dimension
density $\delta$ and the entropy density $\eta$ for such a subsystem
under variation of the length of the subsystem $l$ in Fig. 2. The data
were taken from a CML with $\epsilon = 0.3$ and the total length
$I=100$, starting at $i_0$=40. For the estimation of the Lyapunov
spectrum an observation time $N=20,000$ was taken into account. We
focus on a model with $m=1$. One finds a convergence of the dimension
and the entropy for larger values of $l$, as one expects for extensive
systems.

Furthermore, we estimate Lyapunov spectra of subsystems $\{x_n^i\}
,i=i_0,...,i_0+l-1$ of the SES (\ref{CML}) with a varying spatial
offset $i_0$ and length $l=10$.  With this, we unveil the effects of
boundaries as well as the independence of the spectrum sufficiently
far away from the boundaries. The latter effect can be used to
estimate the spectrum of a large SES from a small subsystem only
\cite{Carretero-Gonzalez98,parekh98} .  The subsystem of length $l=10$
is taken from a CML (\ref{log}) of length ($I=100$) with fixed
boundary conditions and $\epsilon = 0.3$. For the estimation an
observation time $N=20,000$ was taking into account.   We find that the Lyapunov exponents approach limit
values for large enough offset $i_0$, while the values of the
exponents are significantly decreased in the vicinity of the boundary.
From the spectra the dimension density $\delta=D_{KY}/l$ and the
entropy density $\eta=h/l$, where $h$ is the metric entropy, are
estimated as shown in Fig. 3.  For $i_0 > 5$ we observe a plateau of
the dimension and the entropy in the limits of the statistical errors
of the estimations. For $i_0 \leq 5$, both quantities are decreased as
an effect of the fixed boundaries of the system.
While the dynamics as
well as the input sequence at the boundaries of all the subsystems are
different for different spatial offsets $i_0$, we find a 'universal'
Lyapunov spectrum sufficiently far away from the boundaries.  We take
the latter results as an evidence that the Lyapunov spectra of the
non-autonomous subsystems are independent of the external driving at
their boundaries in this case.  Therefore, the Lyapunov-spectra of a
subsystem of a homogeneous SES, sufficiently far away from the
boundaries only depends on the size of the subsystem and is an
appropriate tool to characterize the dynamics of the subsystem as well
as the larger system itself as has been conjectured by Grassberger
\cite{Grassberger89}.

Until now, we described the technique to estimate Lyapunov spectra of
SES from space-time data on a rather fundamental level. In the
following we shortly comment on more general cases. A more profound
discussion will be presented in a subsequent paper. The next step is
to also consider CMLs in 2 and 3 spatial dimensions: The increase in
spatial dimensions immediately implies an increase of the number of
nearest neighbors. Additionally, the structure of the Jacobian will be
such that there are several diagonals with non-zero entries, which are
separated by diagonal bands of zeros.  Also the investigation of
non-homogeneous systems must be taken into account. In this case the
model in eq. (\ref{CML}) depends on the space, $\hat{h}=\hat{h}_i$, as
well as the partial derivatives of the model in
eq. (\ref{Jacobian}). A non-homogeneous model $\hat{h}_i$ does not
introduce any conceptual problems, but a larger number of data is
required since the average over space and time for the fitting of the
model as applied in our example, cannot be applied for a
non-homogeneous model.  For practical purposes it is by far more
interesting to estimate the Lyapunov spectrum of SES continuous in
space and time as modelled by PDEs. The identification of PDEs from
experimental data involves the estimation of derivatives with respect
to space and time as shown in \cite{Voss98,Baer98}.  For the estimation of
the Lyapunov spectrum though, one has to introduce some appropriate
discretization in space and time, as it is common in the case of the
estimation of Lyapunov spectra from a PDE directly. After the
discretization, the procedure evolves along the same lines as
described for CMLs, where the model $\hat{h}$ of the time-evolution
equation (\ref{CML}) takes the form of a discretized PDE, while an
additional dependence of the Lyapunov spectra on the discretization in
space and in time has to be taken into account.  In the case of a
$N$-component CML or PDE with $N>2$ (the state space dimension is
$N$), the measurement of $N$-component space-time data is required for
the estimation of the Lyapunov spectrum.  The important problem of how
to identify and characterize a $N$-component SES from a scalar
space-time measurement is subject to current research.

We acknowledge helpful discussions with M. Abel, A. Giaquinta,
H. Kantz, J. Parisi, A. Politi, and H. Voss.  M. J. B. is supported by
a Marie-Curie-Fellowship of the EU with the contract number:
ERBFMBICT972305. R. H. is partly supported by the EU with the contract
number: ERBFMRCT96.0010 and wants to thank the colleagues at the INO
for their kind hospitality.



\begin{figure}
\psfig{file=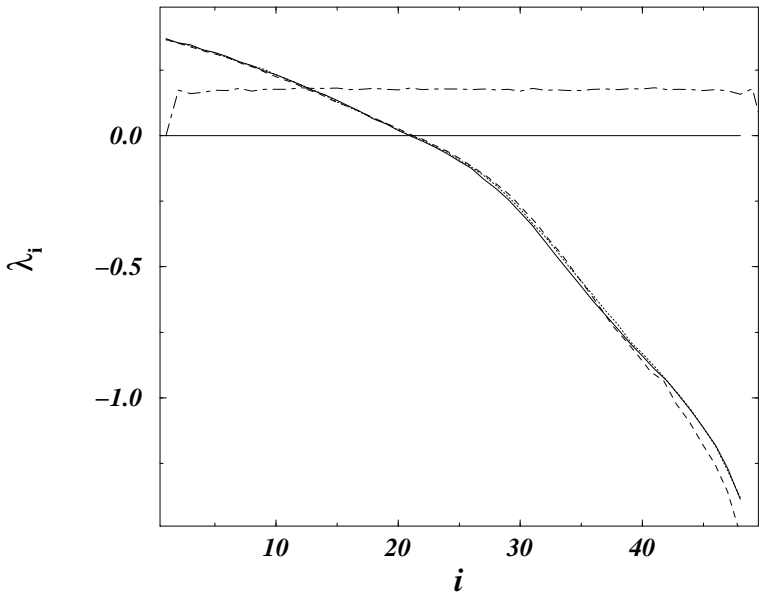,width=10cm,angle=270}
\caption{Lyapunov spectrum of a coupled map lattice of 
50 sites estimated from the equations (solid line), 
and from space-time data with $m=0$ (dot-dashed line;
divided by 10 for presentational purposes), 
$m=1$ (dotted line), and $m=2$ (dashed line).}
\end{figure} 

\begin{figure}
\psfig{file=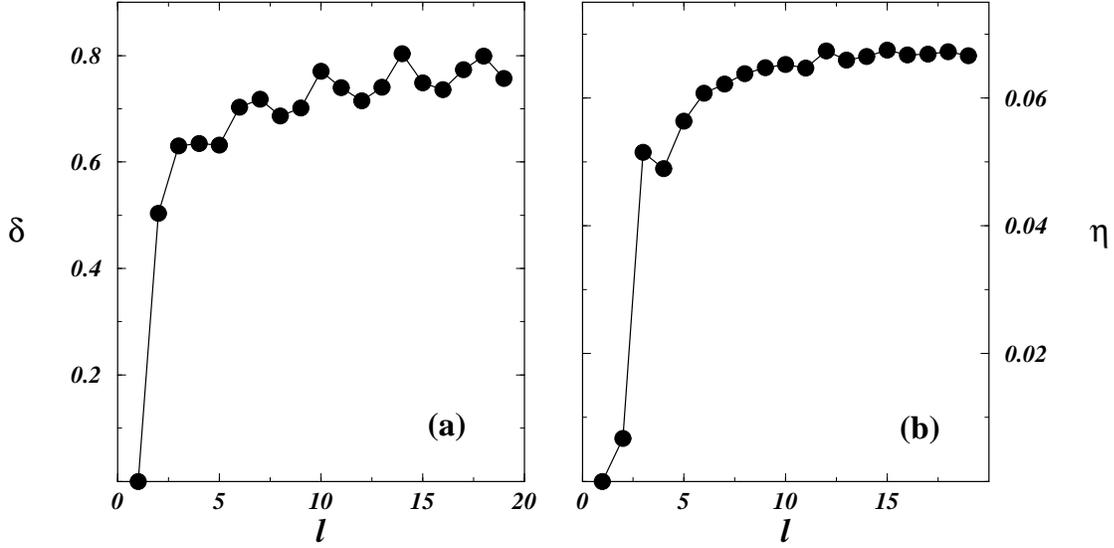,width=15cm,angle=270}
\caption{(a) Dimension density $\delta$, and (b) entropy density $\eta$ 
as estimated from a system of total length $I=100$ (at the site 
$i_0=40$) with varying length $l$ of the subsystem.}
\end{figure} 

\begin{figure}
\psfig{file=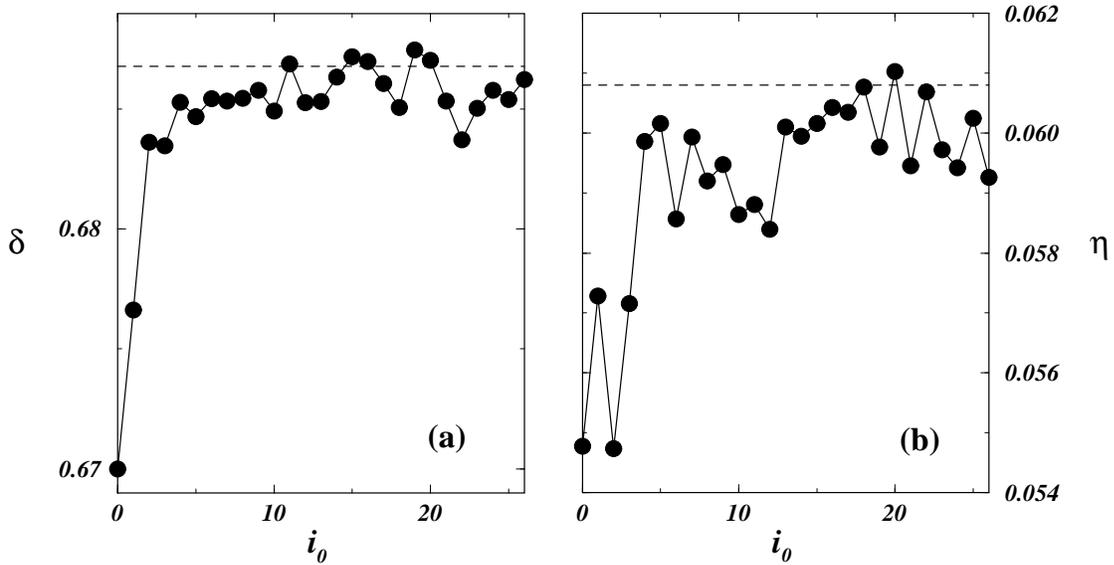,width=15cm,angle=270}
\caption{(a) Dimension density $\delta$, and (b) entropy density $\eta$ 
as estimated from a subsystem of length $l=10$ with varying
distance $i_0$ from the boundaries. The horizontal lines are the
values for the densities taken at $l=8$ according to Fig.~2.}
\end{figure} 

\newpage

\begin{table}
\begin{center}
\begin{tabular}{|l|c|c|c|} \hline
 		& $\delta$ 	& $\eta$ 	& $\lambda_{max}$ \\ \hline
equation 	& 0.72	& 0.085	& 0.37 \\ \hline
data $(m=0)$	& --	& 3.4	& 3.6 \\ \hline
data $(m=1)$	& 0.72	& 0.085	& 0.37 \\ \hline
data $(m=2)$	& 0.70	& 0.085	& 0.37 \\ \hline
\end{tabular}
\end{center}
\caption{Estimation of the dimension density $\delta$, the entropy
density $\eta$, and the largest Lyapunov exponent $\lambda_{max}$ from
the equations (first row), and the from the data (second row).}
\end{table}

\end{document}